\documentclass[12pt]{article}

\usepackage{graphicx}
\usepackage{amsfonts}
\usepackage{amsthm}
\usepackage{amsmath}
\usepackage[top=3cm, bottom=3.25cm, left=2.75cm, right=2.75cm]{geometry}
\newcommand{\Section}[1]{\section{#1} \setcounter{equation}{0}}
\newcommand{\beq}{\begin{equation}}
\newcommand{\eeq}[1]{\label{#1}\end{equation}}
\newcommand{\ber}{\begin{eqnarray}}
\newcommand{\eer}[1]{\label{#1}\end{eqnarray}}
\newcommand{\re}[1]{(\ref{#1})}

\newcommand{\half}{\textstyle\frac 1 2}

\newcommand{\bbD}[1]{\mathbb{D}_{#1}}
\newcommand{\bbDB}[1]{\bar{\mathbb{D}}_{#1}}

\newcommand{\bbX}[1]{\mathbb{X}_{#1}}
\newcommand{\bbXB}[1]{\bar{\mathbb{X}}_{#1}}

\def\one{1\!\!1}
\def\+{{+\!\!\!+}}
\def\pp{\mbox{\tiny${}_{\stackrel\+ =}$}}

\begin{document}
\renewcommand{\theequation}{\thesection.\arabic{equation}}
\setcounter{page}{0}
\thispagestyle{empty}
\begin{flushright} \small
UUITP-14/10\\
\end{flushright}

\smallskip
\begin{center}
 \LARGE
{\bf  Sigma models with non-commuting complex structures and extended supersymmetry.}\footnote{Talk presented by U.L. at ``30th Winter School on Geometry and Physics''
Srni, Czech Republic January 2010.}
\\[12mm]
 \normalsize
{\bf M.~G\"oteman$^{a}$ and U.~Lindstr\"om$^{a}$}, \\[8mm]
 {\small\it
$^a$Theoretical Physics,\\Department Physics and Astronomy,\\
Uppsala University, \\ Box 516, SE-751 20  Uppsala, Sweden.}
\end{center}
\vspace{10mm} \centerline{\bfseries Abstract} \bigskip

\noindent We discuss additional supersymmetries for $\mathcal{N}=(2,2)$  supersymmetric non-linear sigma models
described by left and right semichiral superfields.

\vfill
\eject
\Section{Introduction}

In this short review we summarize some recent development in understanding extended supersymmetry for two-dimensional supersymmetric sigma models described entirely in terms of semichiral superfields.  In the special case of one left and one right semichiral field we find that off-shell supersymmetry is impossible, but that the model gives an interesting description of neutral (pseudo) hyperk\"ahler geometry if we require pseudo supersymmetry instead. In the general case, we encounter Magri-Morosi concomitants in the conditions for closure of the algebra and find that the geometry has a nice description in terms of Yano $f$-structures on the sum of two copies of the tangent space.

Part of this review is based on a paper written in collaboration with Martin Ro\v cek and Itai Ryb.

\section{Preliminaries}
Most of this section is a review of results from \cite{Gates:1984nk}.

Sigma models are collections of maps $\phi^i, ~i=1,...,d,$ from a space $\Sigma$ to a target space $\cal T$:
\ber
 \phi^i :\Sigma  \to \mathcal T
\eer{1}
subject to equations that derive from an action. We shall be concerned with the case when $\Sigma$ is a two-dimensional $(1,1)$ or $(2,2)$ superspace and the $\phi^i$s are real or complex superfields. For $(1,1)$ supersymmetry with real superfields and $D$-algebra
\ber
D^2_\pm = i\partial_{\pp}~,
\eer{2}
the action is 
\ber
\int_\Sigma d^2\xi d^2\theta D_+\phi^i\left(G_{ij}(\phi)+B_{ij}(\phi )\right)D_-\phi^j~.
\eer{3}
Here $D_\pm$ are the covariant derivatives w.r.t.  the anticommuting spinorial coordinates $\theta^\pm$ and $\partial_{\pp}$ are the derivatives w.r.t. the commuting coordinates $\xi^{\pp}$. The field equations that define the sigma model follow from the action \re{3} and read
\ber
\nabla_+^{(-)}D_-\phi^i=D_+D_-\phi^i+D_+\phi^j\Gamma_{jk}^{(-)~i}D_-\phi^k=0~,
\eer{4}
where the connection is defined to be the Levi-Civita connection w.r.t. the metric $G_{ij}$, plus torsion
\ber
\Gamma_{jk}^{(-)~i}:=\Gamma_{jk}^{(0)~i}-\half G^{il}H_{ljk}~, 
\eer{5}
and $H$ is the $B$-field field-strength. We see that a lot of geometry on $\cal T$ enter these expressions. In particular we find that the target space has to carry a torsionful geometry and that the defining equation \re{4} for the sigma model contains the pullback of the Laplacian in this geometry. In fact, in \cite{Gates:1984nk}, analyzing the sigma model \re{3}, the following possible geometries were found:

\begin{table}[htdp]
\begin{center}
\begin{tabular}{|c|c|c|c|c|c|}
\hline
N=(1,1)&N=(1,1)&N=(2,2)&N=(2,2)& N=(4,4)&N=(4,4)\cr
\hline
G&G,B&G&G,B&G&G,B\cr
\hline
Riemannian& Riem. w. Torsion&K\"ahler& Bi-Hermitean&Hyperk\"ahler&Bi-Hyperherm.\cr
\hline
\end{tabular}
\end{center}
\label{default}
\caption{The relation between the number of supersymmetries and type of target-space geometry.}
\end{table}%
In this analysis the starting point is the $(1,1)$ action \re{3} with $G$ or both $G$ and $B$ non-zero. Additional supersymmetries are non-manifest and follow from the ansatz 
\ber
\delta\phi^i = \epsilon^+D_+\phi^kJ^i_{(+)k}(\phi)+ \epsilon^-D_-\phi^kJ^i_{(-)k}(\phi)~.
\eer{6}
This ansatz is for one additional left and one additional right supersymmetry. There may be more.
For $(2,2)$ supersymmetry one finds the following set of conditions from {\em closure of the algebra}:
\ber\nonumber
&J^2_{(\pm)}=-\one\\[1mm]\nonumber
&{\mathcal N}(J_{(\pm)})=0\\[1mm]
&[J_{(+)},J_{(-)}]\cdot (FE)=0~,
\eer{7}
where the first line says that $J_{(\pm)}$ are almost complex structures and $\cal N$ is the Nijenhuis tensor whose vanishing signals their integrability.
Finally, $(FE)$ is an expression that has the structure of a field equation. The upshot of these conditions is that the algebra closes on the kernel of the commutator of the two complex structures and/or on-shell.

{\em Invariance of the action} gives additional constraints:
\ber\nonumber
&\nabla^{(\pm)}J_{(\pm)}=0\\[1mm]
&J^t_{(\pm)}GJ_{(\pm)}=G
\eer{8}
The first condition requires each complex structure to be covariantly constant with respect to the corresponding torsionful connections. The second condition is hermiticity of the metric with respect to both complex structures (bi-hermitean geometry).

More than one extra left and one extra right supersymmetry introduces a set of left and right complex structures $J^{(A)}_{(\pm)}$ that each obey \re{7}, \re{8} as well as some additional constraints both from the algebra and from invariance of the action. The relevant case here is $(4,4)$ supersymmetry where $A=1,2,3$ and the additional constraint from the algebra is that they obey the algebra of quaternions.
Together with the first condition in \re{7}, we write this as
\ber
J^{(A)}_{(\pm)}J^{(B)}_{(\pm)}= -\delta^{AB}+\epsilon^{ABC}J^{(C)}_{(\pm)}~.
\eer{9}

In both the $(2,2)$ and the $(4,4)$ case, on-shell closure of the algebra signals that a manifest $(2,2)$ formulation, if it exists, will have to contain auxiliary fields
\cite{Lindstrom:2004eh}.
On ker$[J^{(A)}_{(+)},J^{(B)}_{(-)}]$, however, a formulation in terms of $(2,2)$ geometry certainly exists \cite{Gates:1984nk}.

The $(2,2)$  $D$-algebra is 
\ber
\{D_\pm,\bar D_\pm\}=i\partial_{\pp}~,
\eer{10}
and the constrained $(2,2)$ superfields that we shall need are chiral $\phi$, twisted chiral $\chi$ and left and right semichiral $\mathbb{X}_{L/R}$ fields obeying
\ber\nonumber
&&\bar D_\pm \phi=0, \quad D_\pm\bar \phi=0\\[1mm]\nonumber
&&\bar D_+\chi =ÊD_-\chi =0, \quad D_+\bar\chi =Ê\bar D_-\bar\chi =0\\[1mm]\nonumber
&&\bar D_+\mathbb{X}_L=0, \quad D_+\bar{\mathbb{X}}_L=0\\[1mm]
&&\bar D_-\mathbb{X}_R=0, \quad D_-\bar{\mathbb{X}}_R=0
\eer{11}
In terms of these fields, a sigma model on ker$[J_{(+)},J_{(-)}]$ is described by the action
\ber
\int d^2\xi d^2\theta d^2\bar\theta K(\phi, \bar \phi,\chi,\bar \chi)~,
\eer{12}
where the $(2,2)$ supersymmetry is now manifest.
When $K$ is independent of either $\phi$ or $\chi$, $B=0$ and  the geometry is K\"ahler, with $K$ being the K\"ahler potential for the metric $G$.
In general, $K$ will be a (linear) potential both for the metric and the $B$-field. More precisely, since $B$ is a gauge-field, it is better to think of $K$ as a potential for $H$.

The $(4,4)$ models on ker$[J^{(A)}_{(+)},J^{(B)}_{(-)}]$ are described by the same action \re{12}, with an equal number of chiral and twisted chiral fields, plus the following conditions on $K$:
\ber\nonumber
&K_{\phi^i\bar\phi^j}+K_{\chi^i\bar\chi^j}=0~~\\[1mm]
&K_{\phi^i\bar\phi^j}-K_{\phi^j\bar\phi^i}=0~.
\eer{13}

\section{The complement of ker$[J_{(+)},J_{(-)}]$. A first look.}

As mentioned in the previous section,  a $(2,2)$ formulation of the sigma models on the complement (ker$[J_{(+)},J_{(-)}])^\bot$ requires auxiliary fields. One type of $(2,2)$-fields whose $(1,1)$ components include (spinorial) auxiliary fields are the semichiral ones \cite{Buscher:1987uw}. We may thus consider
\ber
\int d^2\xi d^2\theta d^2\bar\theta K(\phi, \bar \phi,\chi,\bar \chi,\mathbb{X}_L,\bar{\mathbb{X}}_L,\mathbb{X}_R,\bar{\mathbb{X}}_R)~,
\eer{14}
where, to get a sensible theory, we need to include an equal number of left and right semichiral fields and the generalized K\"ahler potential $K$ satisfies some regularity conditions \cite{Lindstrom:2005zr}. It is shown in \cite{Lindstrom:2005zr} that (away from singular points) \re{14} indeed gives a complete description of bi-hermitean geometry, or, equivalently, of Generalized K\"ahler geometry \cite{gualtieriPhD}, \cite{Lindstrom:2004iw}, \cite{Lindstrom:2007qf}. In what follows we shall be interested in the complement (ker$[J_{(+)},J_{(-)}])^\bot$ only, so we set the chiral and twisted chiral fields to zero and study
\ber
\int d^2\xi d^2\theta d^2\bar\theta K(\mathbb{X}_L,\bar{\mathbb{X}}_L,\mathbb{X}_R,\bar{\mathbb{X}}_R)~.
\eer{15}
The question we pose is what the last entry in Table 1 looks like from this $(2,2)$ perspective. More precisely, we make an ansatz for additional supersymmetries among the semichiral fields and read off the consequences from closure of the algebra and invariance of the action. Clearly we will find the bi-hypercomplex geometry this way, but there may be additional structure involving the auxiliary fields and we also expect to find conditions on $K$ analogous to \re{13} in the kernel.  Previously the question of $(4,4)$ supersymmetry has been partly addressed, in doubly projective superspace, for $(4,4)$ multiplets containing  $(2,2)$ semichiral and  $(2,2)$ auxiliary fields in \cite{Lindstrom:1994mw}. This corresponds to on-shell closure of the fields in the action \re{15} and complements the present analysis.

In our first analysis \cite{Goteman:2009xb}, we restrict ourselves to one set of semichiral fields. The corresponding target-space is thus four (real) dimensional. 
From a general ansatz for the additional supersymmetries we deduce that off-shell closure of supersymmetry, $\{Q,\bar Q\}=i\partial$, is impossible. However, interestingly, if we instead ask for pseudo supersymmetry\footnote{This makes the full symmetry of the model a twisted supersymmetry \cite{AbouZeid:1999em}.}
, $\{Q,\bar Q\}=-i\partial$, off-shell closure is possible and we elaborate the consequences for the case of linear transformations:
\ber
\delta \bbX{L}&=&
i\bar\epsilon^+\bbDB{+}(
\bbXB{L} +\bbX{R}+\tfrac{1}
{\kappa}\bbXB{R})
+i\kappa\bar\epsilon^-\bbDB{-}\bbX{L}-\tfrac{i}{\kappa}\epsilon^-\bbD{-}\bbX{L},\cr
\delta \bbX{R}&=&
i\bar\epsilon^-\bbDB{-}(
\bbXB{R} -
{(\kappa \bar\kappa-1)}
\bbX{L}+
\tfrac{\kappa \bar\kappa-1}{\bar\kappa}
\bbXB{L})
-i
\bar\kappa \bar\epsilon^+\bbDB{+}\bbX{R} +
\tfrac{i}{\bar\kappa} \epsilon^+\bbD{+}\bbX{R}~.
\eer{16}
The only parameter here, modulo field redefinition and $R$-symmetry, is the complex parameter $\kappa$. The asymmetry between left and right fields is irrelevant and is  an artefact of our choices of redefinitions. These transformations close off-shell to a pseudo supersymmetry. 

Invariance of the action \re{15} under the transformations \re{16} requires that the following equations are fulfilled
\ber
	K_{1\bar 1} - K_{12} - \bar\kappa K_{\bar 1 2} &=& 0,\cr
	(\kappa \bar\kappa -1)K_{2\bar 2} + K_{12} - \kappa K_{1\bar 2} &=& 0~,
\eer{17}
where the subscripts $1$ and $2$ denote derivatives w.r.t. the left and right semichiral field, respectively. The system (\ref{17}) may be solved by separating  variables to give a two-parameter family of solutions
\ber
K=F(y) + \bar F(\bar y), \quad y= \alpha \bbX{L}+\beta \bbXB{L} +\gamma \bbX{R} +\delta \bbXB{R},
\eer{18}
where
\ber
\gamma = 
\frac{\alpha\beta}{\alpha +\bar\kappa \beta}, \quad
\delta = \frac{\alpha\beta} {\kappa\alpha + \beta}.
\eer{19}
Due to the linearity of the conditions (\ref{17}), the solution integrated over the free parameters is again a solution. The general $K$ is thus
\ber
K(\mathbb{X}_L,\bar{\mathbb{X}}_L,\mathbb{X}_R,\bar{\mathbb{X}}_R)=\int d\alpha d\beta ~{\mathcal K}(\alpha,\beta; \alpha \bbX{L}+\beta \bbXB{L} +\gamma \bbX{R} +\delta \bbXB{R})~,
\eer{20}
where ${\mathcal K}$ is a particular solution of the type \re{18}. 

It is an interesting fact that the complex structures that follow from the solution \re{20} fulfill\footnote{Here we have chosen $\kappa=\sqrt{2}$ for definiteness, the argument goes through for any $\kappa$ and only depends on the fact that $|\frac{\kappa \bar\kappa+1}{\kappa \bar\kappa-1}|>1$.}
\ber
\{J_{(+)},J_{(-)}\}=-6\cdot\one~.
\eer{21}
This means that the $H$-field is trivial \cite{Lindstrom:2005zr} and also that there are two local product structures
\ber
S:=\frac 1 {2\sqrt{2}} \left(J_{(+)}-3J_{(-)}\right)~, \quad T:=\frac 1 {4\sqrt{2}} [J_{(+)},J_{(-)}]~,
\eer{22}
that preserve the metric $G$ of signature $(2,2)$, and together with $J_{(+)}$ generate $SL(2,\mathbb{C})$. The corresponding geometry is called  neutral (pseudo) hyperk\"ahler.
We have thus reached an interesting conclusion; starting from a $(2,2)$ sigma-model described by a generalized K\"ahler potential $K$, the requirement that it in addition carries non-manifest pseudo supersymmetry leads to neutral hyperk\"ahler geometry on the target-space. Since $K$ is a potential for all geometric objects in the $(2,2)$ model, including metric and complex structures, (albeit entering in a non-linear way), we see that it is also a potential for all the geometric objects in the neutral hyperk\"ahler geometry. Furthermore, our approach provides a recipe for constructing such geometries.  
\section{The general case.}
In this section we give a brief summary of the results of \cite{Goteman:2009ye}.
When there are more then one set of semichiral fields, there is no problem with off-shell (ordinary) supersymmetry. The general ansatz is
\ber\nonumber
\bar\delta^{(\pm)}\mathbb{X}=U^{(\pm)}\bar \epsilon^\pm\bar D_\pm\mathbb{X}~\\[1mm]
\delta^{(\pm)}\mathbb{X}=V^{(\pm)}\epsilon^\pm D_\pm\mathbb{X}~,
\eer{23}
where
\ber
\mathbb{X}:=\left(\begin{array}{c}
\mathbb{X}_L\cr
\bar{\mathbb{X}}_L\cr
\mathbb{X}_R\cr
\bar{\mathbb{X}}_R
\end{array}\right)=:\left({\mathbb{X}}^i\right)~,
\eer{24}
and where we have assumed that there are an equal number $d$ of left and right fields but have suppressed the corresponding index, displaying only the general index $i=1,...,4d$. The matrices $U$ and $V$ are related via complex conjugation and rearrangement of the rows and columns. Note that the chirality constraints \re{11} imply that not all the entries in $U$ and $V$ are determined, in particular they may be degenerate.

{\em Closure of the algebra} gives the following equations:
\ber
[ \bar{\delta}^\pm,\bar{\delta}^\mp ]\mathbb{X}^i = 0 & 
\Longleftrightarrow &
{\mathcal{M}}(U^{(\pm)},U^{(\mp)})^i_{jk} 
\bar{\mathbb{D}}_{\pm}\mathbb{X}^j 
\bar{\mathbb{D}}_{\mp}\mathbb{X}^k  =
[U^{(\pm)},U^{(\mp)}]^i{}_m\bbDB{\pm} \bbDB{\mp}\bbX{m}~,\nonumber \\
{}[ \bar{\delta}^\pm,{\delta}^\mp ]\mathbb{X}^i = 0 & \Longleftrightarrow &
{\mathcal{M}}(U^{(\pm)},V^{(\mp)})^i_{jk} 
\bar{\mathbb{D}}_{\pm}\mathbb{X}^j{\mathbb{D}}_{\mp}
\mathbb{X}^k  =[U^{(\pm)},V^{(\mp)}]^i{}_m\bbDB{\pm} \bbD{\mp}\bbX{m}~,\nonumber\\{}
[ \bar{\delta}^\pm,\bar{\delta}^\pm ]\mathbb{X}^i = 0 &
\Longleftrightarrow &\mathcal{N}(U^{(\pm)})^i_{jk}\bar{\mathbb{D}}_{\pm}\mathbb{X}^j\bar{\mathbb{D}}_{\pm}\mathbb{X}^k = 0~,
\eer{25}
and
\ber\nonumber
&&[\delta^\pm,\bar{\delta}^\pm]\mathbb{X}^i= i \bar{\epsilon}^\pm \epsilon^\pm \partial_{\pp}\mathbb{X}^i   \Longleftrightarrow \\{}
&&\mathcal{M}(U^{(\pm)},V^{(\pm)})^i_{jk} 
\bar{\mathbb{D}}_{\pm}\mathbb{X}^j \mathbb{D}_{\pm}\mathbb{X}^k
=\left[\left(UV\right)^{(\pm)i}_{~j}+\delta^i_j
\right]\bbDB{\pm}\bbD{\pm}\bbX{}
 +\left[\left(VU\right)^{(\pm)i}_{~j}+\delta^i_j\right]\bbD{\pm}\bbDB{\pm}\bbX{j}~.\nonumber \\{}
\eer{252}
Here $\cal{N}$ is the Nijenhuis tensor  previously mentioned, and $\cal{M}$ is the Magri-Morosi concomitant. Whereas the vanishing of $\mathcal{N}(J)$ for an almost complex structure $J$ implies that it is integrable, vanishing of ${\cal{M}}(J,L)$ for two commuting almost complex structures is related to their simultaneous integrability.
In fact, ${\cal{M}}(J,L)$  is defined for two arbitrary endomorphisms $J$ and $L$, but is only a tensor when they commute.

Off-shell we have no relations between second derivatives and the product of two derivatives. The relations \re{25}-\re{252} can thus only be satisfied if the left and right sides vanish independently. This sets the Nijenhuis tensors, the concomitants and the commutators to zero separately (in the non-vanishing directions). There are many interesting aspects of this off-shell geometry. Here we only mention one: the existence of a Yano $f$-structure \cite{Yano:1961} on $TM\oplus TM$.

Consider the following $4d\times 4d$ matrix defined on the sum of two copies of the tangent space:
\ber
\mathcal{F}_{(\pm)}:=\left(\begin{array}{cc}0&U^{(\pm)}\cr
V^{(\pm)}&0
\end{array}\right)~.
\eer{26}
The entries in $U$ and $V$ that are left undetermined in their definitions are now set to zero. Due to some of the conditions in \re{252}, $\mathcal{F}_{(\pm)}$ satisfies the condition for a Yano $f$-structure
\ber
\mathcal{F}_{(\pm)}^3+\mathcal{F}_{(\pm)}=0~,
\eer{27}
which is slightly weaker that the condition for an almost complex structure in that it allows for degenerate matrices.

Further, the two distributions that correspond to $-\mathcal{F}_{(\pm)}^2$ and $1+\mathcal{F}_{(\pm)}^2$ are integrable in the sense of Yano, due to the vanishing of the Nijenhuis-tensors in  \re{25}.

Now, {\em invariance of the action} \re{15} gives the following set of differential equations
\ber
\left(K_{i}U^{(+)i}{}_{[j}\right){}_{k]}=0, \quad j,k \neq a,
\eer{28}
where $a$ represents the left chiral directions. Similar relations hold for $U^{(-)}$ and $V^{(\pm)}$. These relations are the semichiral counterpart of \re{13} for commuting complex structures and corresponds to \re{17} in the pseudo supersymmetric model.

It is interesting that \re{28} has a geometric formulation related to the $f$-structure, at least when $U$ and $V$ are curl-free in the lower indices. It may then be written
as a suitable projection of
\ber
\mathcal{F}_{(\pm)}^t\mathfrak{B}\mathcal{F}_{(\pm)}=\mathfrak{B}~,
\eer{29}
where we have combined the Hessian $K_{ij}$ of the generalized K\"ahler potential into an
antisymmetric tensor $\mathfrak{B}$ on $TM\oplus TM$ as 
\ber
 \mathfrak{B} = \left(\begin{array}{cc}
0 & K \\ -K^t & 0
\end{array}\right).
\eer{30}

\section{On-shell}

We know from \cite{Gates:1984nk} that the underlying geometry for the $(4,4)$ theory is bi-hypercomplex, also on (ker$[J_{(+)},J_{(-)}])^\bot$ .
We may compare our results to this by imposing the field equations. In a real basis for the covariant derivatives,
\ber
D_\pm =:\half (\mathcal{D}_\pm-i\mathcal{Q}_\pm)~,
\eer{31}
going on-shell amounts to setting
\ber
D_\pm\mathbb{X}^i=(\bar \pi^{(\pm)}\mathcal{D}_\pm\mathbb{X})^i~,
\eer{31}
where the projection operator is
\ber
\pi^{(\pm)}:=\half \left(1+iJ_{(\pm)}\right)~.
\eer{32}
Identifying $J_{(\pm)}$ with $J_{(\pm)}^{(3)}$ of the quaternion worth of complex structures in the bi-hypercomplex geometry, and identifying the transformations
\re{23} with the known transformations in terms of $J_{(\pm)}^{(1)}$ and $J_{(\pm)}^{(2)}$ on shell, we identify
\ber\nonumber
&\half\left(J_{(\pm)}^{(1)}-iJ_{(\pm)}^{(2)}\right)=U^{(\pm)}\pi\\[1mm]
&\half\left(J_{(\pm)}^{(1)}+iJ_{(\pm)}^{(2)}\right)=V^{(\pm)}\bar\pi
\eer{33}
on shell.

Since the on-shell conditions gives a  relation between second derivatives and the product of two derivatives, the relations in \re{25} have more solutions.
Clearly the off-shell transformations form a subset of the on-shell ones. In fact, with the identification \re{33} all the relations in \re{25} should be satisfied, possibly up to constraints from invariance of the action, since the bi-hypercomplex geometry gives the full answer. We have checked that this is indeed the case.
\bigskip

{\bf Acknowledgements:} We are grateful to our collaborators on one of the papers reviewed, Martin Ro\v cek and Itai Ryb . The research of UL was supported by VR grant 621-2009-4066.

\end{document}